\documentclass[aps,pra,twocolumn,reprint]{revtex4-1}
\usepackage{amsmath,amssymb,graphicx,hyperref}
\newcommand{\eqn}[1]{\begin{eqnarray} #1 \end{eqnarray}}
\newcommand {\LL} {\perp}
\newcommand {\tbf}[1] {\textbf{#1}}
\newcommand {\tit}[1] {\textit{#1}}
\newcommand {\trm}[1] {\textrm{#1}}
\newcommand {\scr}[1] {\mathcal{#1}}

\newcounter{defn}
\newcommand{\defn}[1]{\refstepcounter{defn}\label{#1}}
\newcounter{lemm}

\newcounter{thrm}
\newcommand{\thrm}[1]{\refstepcounter{thrm}\label{#1}}

\begin{document}

%\title{Which causal scenarios are interesting?}
\title{Which causal structures might support a quantum-classical gap?}
\author{Jacques Pienaar}
 \email{jacques.pienaar@zoho.com}
 \affiliation{
 Faculty of Physics, University of Vienna, Boltzmanngasse 5, A-1090 Vienna, Austria.
}
 \affiliation{
 Institute of Quantum Optics and Quantum Information,
Austrian Academy of Sciences, Boltzmanngasse 3, A-1090 Vienna, Austria.
}

%\tableofcontents

\begin{abstract}
{A causal scenario is a graph that describes the cause and effect relationships between all relevant variables in an experiment. A scenario is deemed `not interesting' if there is no device-independent way to distinguish the predictions of classical physics from any generalised probabilistic theory (including quantum mechanics). Conversely, an interesting scenario is one in which there exists a gap between the predictions of different operational probabilistic theories, as occurs for example in Bell-type experiments. Henson, Lal and Pusey (HLP) recently proposed a sufficient condition for a causal scenario to not be interesting. In this paper we supplement their analysis with some new techniques and results. We first show that existing graphical techniques due to Evans can be used to confirm by inspection that many graphs are interesting without having to explicitly search for inequality violations. For three exceptional cases -- the graphs numbered $\# 15,16,20$ in HLP -- we show that there exist non-Shannon type entropic inequalities that imply these graphs are interesting. In doing so, we find that existing methods of entropic inequalities can be greatly enhanced by conditioning on the specific values of certain variables.}

\end{abstract}

\maketitle

\section{Introduction}

Every physical experiment has an underlying causal structure, conceived of as a series of \tit{events}, and the \tit{influences} they have on each other. Just as trees look much the same when stripped of their leaves, it is possible for experimental set-ups in widely differing contexts to nevertheless exhibit the same causal structure. Causal structure is a useful abstraction that permits us to classify experiments by their cause-and-effect relationships, ignoring the unnecessary details of their physical implementation.

\begin{figure}[!htbp]
\includegraphics[width=8cm]{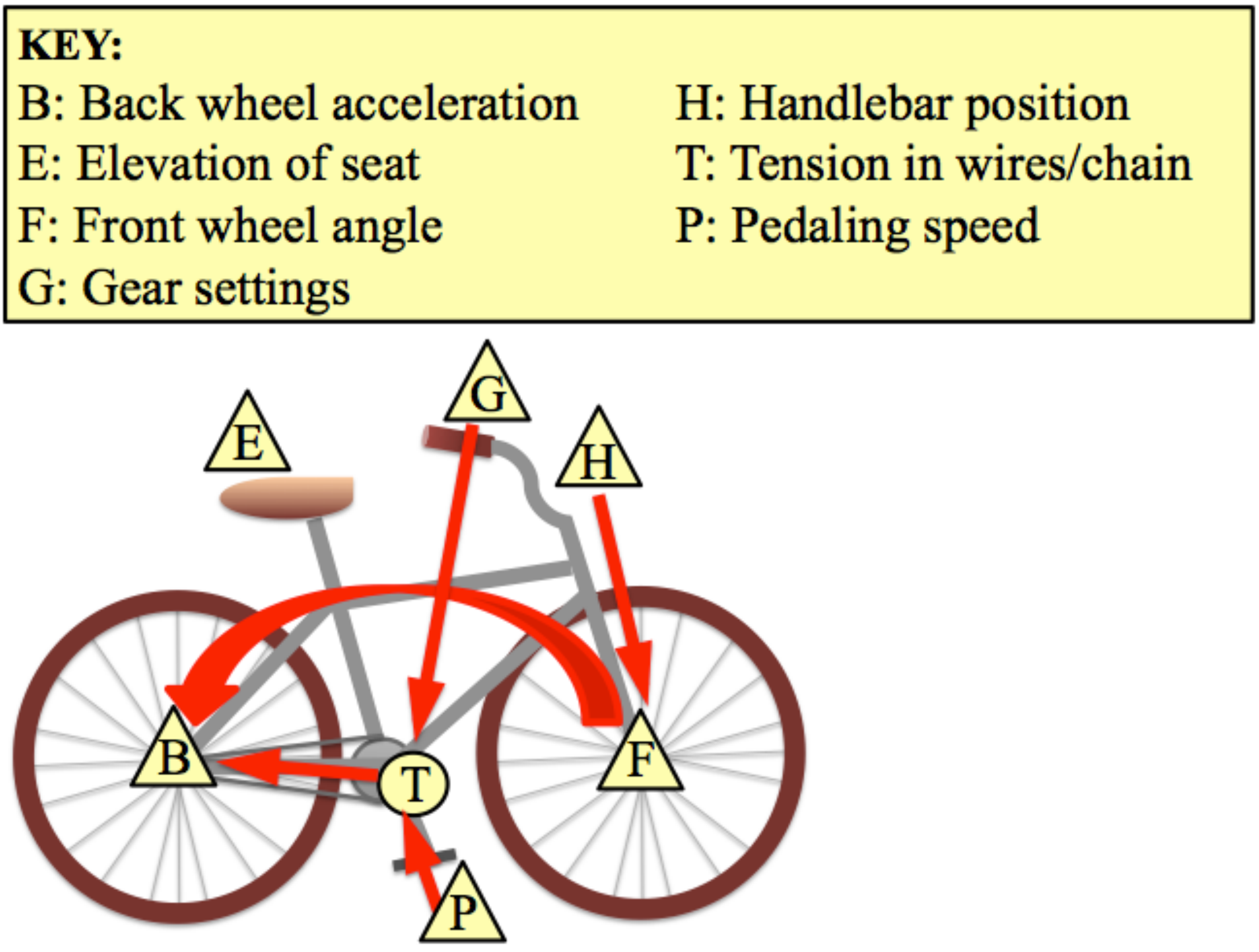}
\caption{A causal graph describing the functioning of a bicycle. The gear settings $G$ (including brakes) and the pedalling speed $P$ influence the acceleration vector $B$ of the back wheel, via the tension $T$ in the wire cables and in the bicycle chain. Since the tension is not directly visible to the eye, we consider it a latent variable (hence it is represented by a circular node). The handlebar position $H$ influences the direction of $B$ indirectly, via its influence on the front wheel orientation $F$. The direct influence of $F$ on $B$ is made possible by the frame of the bicycle. The seat elevation $E$ is causally unrelated to the other variables, so is represented by an isolated node (such nodes are usually omitted from the graph as irrelevant).}
\label{fig:ex1}
\end{figure}

In this spirit, even a complicated physical system has a simple representation in terms of a directed acyclic graph (DAG) describing its causal structure, like that of the bicycle shown in Fig \ref{fig:ex1}. In this graph (called the \tit{causal graph}), the triangular nodes represent the minimal set of observed events that are needed to describe the outcomes of experimental probing of the bicycle. Circular nodes represent \tit{unobserved variables}, which are not directly measured but can affect the observed outcomes. The arrows indicate direct causal influences between events. The probabilities for the values of observed variables are governed by a set of functions relating each variable to its direct causes in the graph; these functions are called the \tit{model parameters}. A causal graph dressed with a set of model parameters is called a \tit{causal model} \cite{PEARL, SGS}.

Following Ref. \cite{HLP}, a causal graph containing both observed and unobserved nodes is called a \tit{generalized} DAG (GDAG). In general, the unobserved variables in a GDAG can represent other objects besides random variables, such as quantum states or even `black boxes' from some hypothetical theory beyond quantum mechanics. Traditionally, however, the unobserved nodes are just random variables whose values are marginalised (summed over), in which case they are called \tit{latent variables}. In this case, the causal model is called \tit{classical}.

What are causal models good for? A typical scientific experiment consists of observing some phenomenon in the laboratory under controlled and repeatable conditions, and then asking whether the probability distribution of the observed events has an explanation in terms of some known physical theory. A causal model provides us with a formal definition of \tit{explanation}. If we can find a causal model that reproduces the observed probabilities, then we say that the observed phenomenon can be explained by the model. Hypothesis testing then becomes the work of designing experiments to rule out different competing explanations.

The literature on \tit{causal inference} provides numerous tools for deciding when one causal model is a better explanation than another. In general, simpler explanations are preferable to more complex ones. Explanations should be \tit{faithful}, meaning (roughly) that the observed independencies appear in almost all causal models that have the same causal structure as the chosen explanation. If these principles are not enough to differentiate different hypotheses, it is always possible to do so by performing experimental \tit{interventions}: actively changing the values of some variables and observing the reaction of the remaining variables. A tool of causal inference called the \tit{do-calculus} tells us which interventions are needed to obtain specific information about the causal structure.

The question arises as to whether certain experiments in quantum physics, particularly the so-called `Bell-type' experiments \cite{BELL76,CHSH,BELLREV}, can be explained by a causal model. It is widely agreed that any explanation of quantum phenomena in terms of a classical causal model must violate at least one of several assumptions that hold for purely classical phenomena. For example, a classical causal model can explain quantum phenomena if we allow causes to propagate faster than light, as in Bohmian mechanics \cite{BOHM}, or if we allow measurement settings to be influenced by a hidden common cause (the super-determinism loophole), and there are many other options. A more recent perspective due to Wood \& Spekkens is that classical causal explanations of quantum experiments cannot be \tit{faithful} \cite{WOOD}, i.e. the causal model is not a generic example of the set of models that share its causal structure. 

An alternative to the above options is to reject classical causal models as explanations for quantum phenomena, and to seek a more general notion of a \tit{quantum} causal model. Such a model would ideally provide a faithful account of quantum phenomena and enable causal inference in the quantum domain \cite{RIED, LEIF13, COST, PIE15, ALLEN}. 

Going a step further, Henson, Lal and Pusey (HLP) \cite{HLP} have shown that it is possible to define yet more general causal models than quantum causal models. One can consider a causal explanation in terms of any hypothetical \tit{generalised probabilistic theory} (GPT), of which classical and quantum causal models are just special cases. The key to the generalisation lies in the way we interpret unobserved nodes. For a classical causal model, an unobserved node is a latent variable; in a quantum causal model, it represents a quantum resource, such as an entangled state; more generally, it represents a `black box' resource of some GPT.

\begin{figure}[!htbp]
\includegraphics[width=8cm]{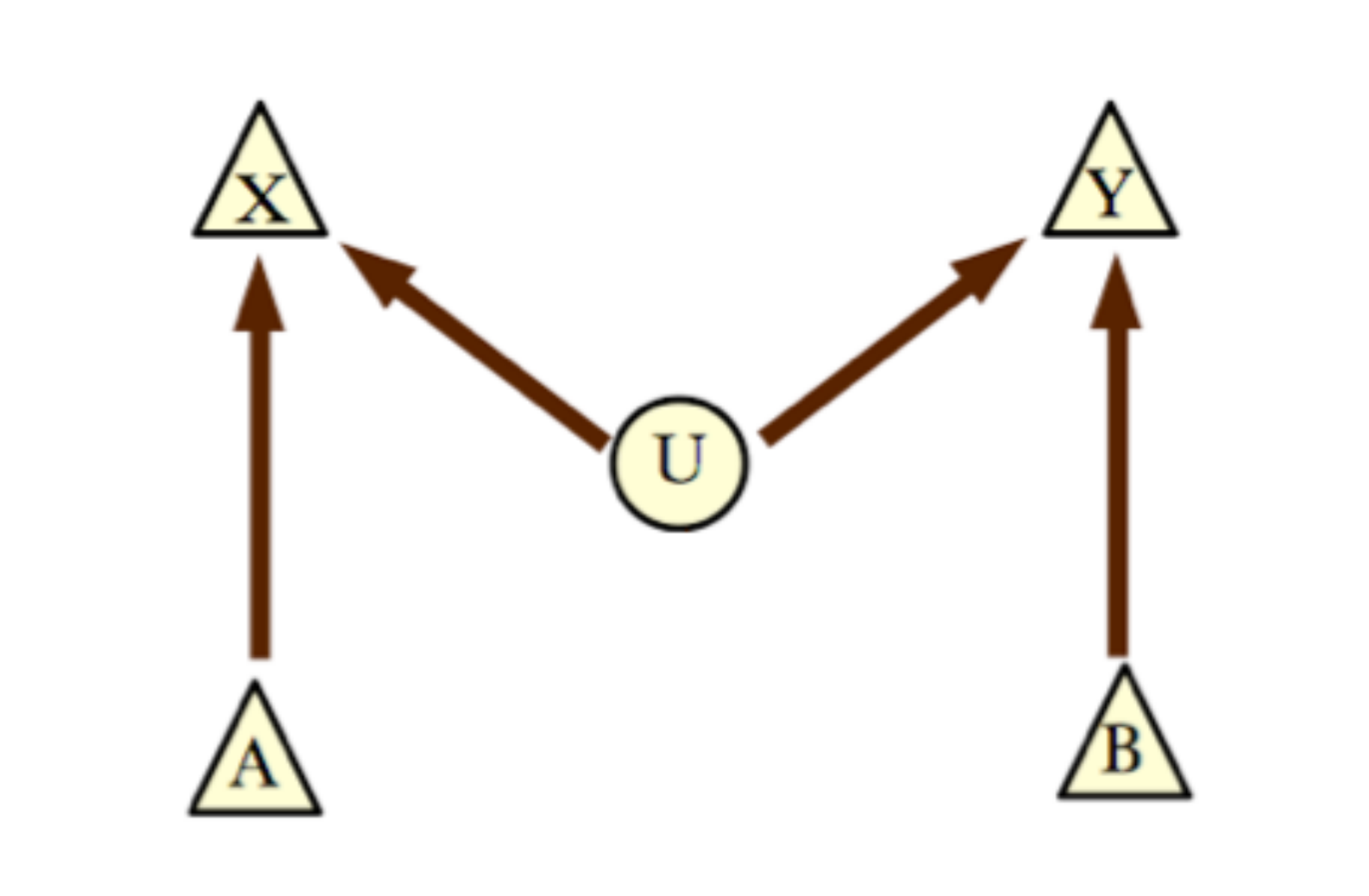}
\caption{A causal graph describing a generic Bell-type experiment, in which a single source produces pairs of particles that are sent to a pair of detectors. $A$ and $B$ are the settings of measurements on each particle, and $X,Y$ are the respective outcomes. These outcomes are influenced by the unobserved node $U$, typically interpreted as the state of the physical particles produced by the source. This state could be a latent random variable in the classical case (like the state of the bicycle chain in Fig. \ref{fig:ex1} \,), but it could also represent a quantum state or a generalized probabilistic resource like a `PR-box'~\cite{HLP}.}
\label{fig:Bell}
\end{figure}

The recognition that different laws of physics necessitate different frameworks for causal inference leads to the conclusion that what constitutes an \tit{explanation} depends on what model of the physical world we choose to take as primitive. It is especially interesting to ask whether, in an experiment whose causal structure is given, there exist differences between the probability distributions that can be explained by different physical models. An example is the causal structure of Bell-type experiments, shown in Fig. \ref{fig:Bell}: if this GDAG is dressed with classical model parameters, the probability distributions satisfy \tit{Bell inequalities}; if we use quantum model parameters, the distributions satisfy the weaker \tit{Tsirelson's bound}\cite{TSIR}; and if we use a GPT that allows \tit{Popescu-Rohrlich boxes}\cite{PRBOX}, the only relevant bound is the no-signalling constraint. 

The fact that this same causal graph supports different sets of observed probabilities means that experiments within this causal structure can be used to perform inference \tit{on the laws of physics themselves}. In effect, by assuming that the causal structure has a given form, we are able to distinguish different physical theories within that structure. This is, after all, precisely why Bell-type experiments are commonly interpreted as ruling out classical physics (i.e. a faithful classical causal model) in favour of quantum mechanics. If a resource were ever discovered that could violate Tsirelson's bound in a Bell-type experiment, this would in turn constitute evidence against the general validity of quantum mechanics. For this reason, the causal graph of this experiment (called the \tit{Bell scenario}) is considered to be `interesting'.

\begin{figure}[!htbp]
\includegraphics[width=8cm]{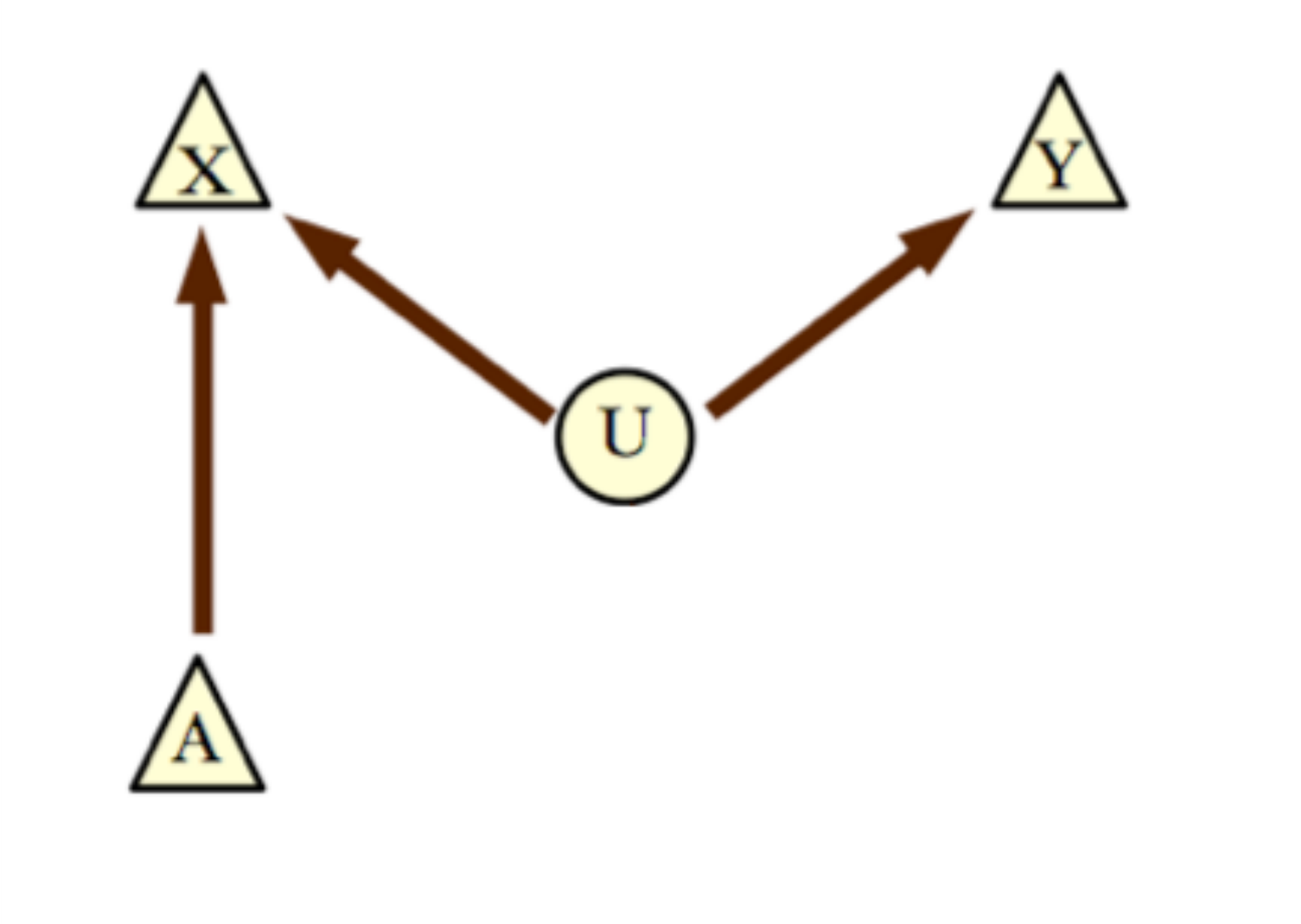}
\caption{A causal graph describing a `one-sided' Bell-type experiment. Only one measurement setting $A$ is freely chosen; the other setting is fixed to a particular value and so is not shown in the graph. The graph is \tit{uninteresting}, because it allows a classical explanation regardless of the underlying physical theory. This is also true for variations of the scenario in which the value of $B$ changes in each run but is not `freely chosen'; see for example Fig.26(a) in Ref.\cite{WOOD}.}
\label{fig:OneSide}
\end{figure}

Following this intuition, HLP gave a precise meaning to the term \tit{interesting} as a formal property of a causal graph (we review their definition in the next section). Not every causal graph is interesting in the sense of HLP's definition. Trivial examples include DAGs in which all variables are observed (these always have a classical explanation), or GDAGs in which all variables share a single common cause (an example of \tit{superdeterminism}). A less trivial example is the so-called `one-sided Bell scenario' (see Fig. \ref{fig:OneSide}), in which one of the measurements is fixed (hence not a random variable). In this case, although there is a non-trivial no-signalling constraint imposed by the causal structure ($A$ cannot send signals to $Y$), it turns out that all possible observed probability distributions can be faithfully reproduced by some choice of classical model parameters. We refer to causal structures like this one as \tit{uninteresting}, because experiments based on them have no hope of differentiating between classical, quantum or supra-quantum theories.

The question now arises how to decide whether a given causal graph is uninteresting or interesting. In their paper, HLP introduced a sufficient criterion for a graph to be uninteresting, and conjectured its necessity as well. To provide evidence for this conjecture, they tried to show explicitly that graphs failing their criterion must be interesting, i.e. must allow probability distributions not achievable by any classical causal model. They were able to do this for almost all graphs of up to six nodes. Only three graphs, labelled \# 15,16,20 (see Fig. \ref{fig:musketeers}), resisted both the sufficient criterion for un-interestingness \tit{and} the authors' attempts to find any interesting probability distributions on the graphs. A main result of the present paper is to resolve this impasse.

The method used by HLP to demonstrate interestingness is somewhat tedious, as it requires identifying an explicit constraint (eg. an entropic constraint or algebraic inequality) and then showing that the constraint does not follow from the causal structure alone. Often the best way to achieve this is to find a counterexample -- a distribution that violates the constraint in question while satisfying all the causal constraints -- but this requires performing an explicit search of the probability space. In this paper, we show that this tedious process can sometimes be sidestepped using known techniques from the literature, which allow interestingness to be ascertained just by inspecting the graph.

Unfortunately, these techniques do not help us in the case of graphs \# 15,16,20, and we are ultimately forced to derive a new kind of inequality for these graphs, which can then be violated by an appropriately defined probability distribution in order to establish their interestingness. Although tedious, this method reveals a new class of \tit{fine-grained} inequalities of non-Shannon type. These inequalities show some promise for being generalized to more complicated scenarios.

The paper is organised as follows. Sec. \ref{Sec:formalism} contains essential notation and background. We then show in Sec. \ref{Sec:skelly} that some simple graphical methods due to Evans\cite{EVANS12,EVANS14} can be used to confirm the interestingness of a graph by inspection. To illustrate the usefulness of this method, we apply it to the graphs considered by HLP and confirm their results. The only cases that do not satisfy Evan's sufficient criteria for interestingness are the Bell and Triangle scenarios, and the graphs \# 15,16,21. For the latter three, in Sec. \ref{Sec:newstuff} we propose a novel extension of the methods of entropic inequalities \cite{CHAV14,CHAV15,CLG14} that confirms these graphs as interesting (and thereby completes the establishment of HLP's conjecture for graphs with up to six nodes). Sec. \ref{Sec:ending} contains our conclusions and outlook.

\section{Definitions and statement of the problem \label{Sec:formalism}}

In this section we review notation and definitions that will be used in this paper. We refer the reader to Refs. \cite{PEARL, SGS} for background on causal models, and to Refs. \cite{CHAV14, CHAV15, CLG14, WEIL} for background on entropic bounds for causal graphs relevant to this paper. 

In this work, we are concerned with joint probability distributions $P(A,B,C...)$ of random variables denoted by capital Roman letters $A,B,C, ...$. Sets of random variables will also be denoted by capital Roman letters; the context will make it clear whether a letter refers to a single variable or a set of variables. For expressions involving set unions, we omit the `$\cup$', for example $XY := X \cup Y$ is understood. Each variable takes values in a discrete, finite set. Whenever we need to make specific reference to them, we take these values to be positive integers. When talking about a specific joint probability distribution, eg. $P(X,Y,Z)$, we will often represent the marginal distributions of $P$ simply by omitting the relevant variables from $P$'s domain; eg. $P(X) := \sum_{Y,Z} \, P(X,Y,Z)$. A \tit{conditional probability}, denoted $P(X|Y)$ for disjoint sets of variables $X,Y$, defines a family of probability distributions on the domain of $X$, parametrized by the discrete index $Y$. Thus, for example, $P(X|Y=1)$ and $P(X|Y=3)$ are two different distributions of the variable $X$, both belonging to the parametrized family of distributions $P(X|Y)$. The law of total probability relates all these concepts in a simple equation: $P(X,Y)=P(X|Y) P(Y)$. 

There are two key properties of joint probability distributions that can be used to characterise them: the \tit{conditional independence (CI)} relations between variables, and the \tit{entropic} relations between the variables. We discuss each of these in turn. \\

\tbf{Definition \defn{def:CI} \ref{def:CI}: CI relations } Let $X, Y, Z$ be three disjoint sets of variables with joint distribution $P(X,Y,Z)$. The sets $X$ and $Y$ are said to be \tit{conditionally independent} given $Z$, denoted $(X \LL Y |Z)$, iff $P(X|Y,Z)=P(X|Z)$.\\

Informally, this means that learning the value of $Y$ provides no new information about the value of $X$ if we already know the value of $Z$. Given a distribution $P(V)$ on a set of variables $V$, the set denoted $\trm{CI}_{P}(V)$ contains all conditional independence relations that hold in $P$ between all disjoint subsets of $V$. Any set of CI relations implies further CI relations via the \tit{semi-graphoid axioms}, which are in turn derivable from the axioms of probability theory. We do not reproduce these here as we will not require them explicitly. A set of CI relations is called \tit{complete} iff it is closed under the semi-graphoid axioms. Unless stated otherwise, all sets of CI relations referred to in this paper are assumed to be complete.

Of particular interest are those complete sets of CI relations that can be \tit{represented} by a DAG, where each node in the DAG represents a variable. Representation means that the CI relations are obtained from the graph using a criterion called \tit{d-separation}:\\

\tbf{Definition \defn{def:dsep} \ref{def:dsep}: d-separation } Given a DAG for a set of variables $V$, two disjoint subsets of variables $X$ and $Y$ are said to be \tit{d-separated} by a third disjoint set $Z$, denoted $(X \LL Y | Z)_d$, iff all undirected paths (i.e. ignoring the direction of arrows) from $X$ to $Y$ are \tit{blocked} by a member of $Z$. A path is blocked by a member of $Z$ iff:\\
(i) the path contains a chain $i \rightarrow m \rightarrow j$ or a fork $i \leftarrow m \rightarrow j$ such that the middle node $m$ is in $Z$; or\\
(ii) the path contains a collider $i \rightarrow m \leftarrow j$ such that the node $m$ and its descendants are not in $Z$.  \\
If a path is not blocked, it is said to be \tit{unblocked}. If there is no path connecting two nodes, as in a disconnected graph, they are trivially d-separated by all other subsets of nodes. \\

By imposing a correspondence between d-separation and CI relations, $(X \LL Y| Z)_d \Rightarrow (X \LL Y | Z)$, every DAG $G(V)$ implies a complete set of CI relations on the variables $V$, denoted $\trm{CI}_{G}(V)$.

The link between CI relations and causal models is as follows: a causal model is regarded as a possible explanation of a distribution $P(V)$ if the set of CI relations $\trm{CI}_{G}(V)$ obtained from the model's DAG are a subset of the set $\trm{CI}_{P}(V)$. More generally, we might regard $P(V)$ as the marginal of an unknown distribution $P(V,U)$ and seek an explanation in terms of a GDAG $G(V,U)$ that includes latent variables $U$. In this case consistency requires that the subset of CI relations $\trm{CI}_{G}(V)$ (i.e. restricted to the observed variables $V$) should be a subset of the CI relations in the marginal $\trm{CI}_{P}(V)$. 

Since our main goal in this work is to classify the possible distributions that can arise from a given GDAG ignoring interventions, we do not review the details of causal inference, although we refer to it in passing in Sec. \ref{Sec:newstuff}.

We now turn to a different property of probability distributions: the \tit{entropic} constraints between variables. Given any set of variables $X$ distributed by $P(X)$, we can associate an \tit{entropy}, $H(X)$. A common example is the \tit{Shannon entropy} \cite{SHAN} defined by:
\eqn{
H(X) := - \sum_{X} \, P(X) \trm{log}_2 \left[ P(X) \right] \, ,
} 
where the sum is over the values of all variables in $X$. Given an entropy function, it is also useful to define the \tit{conditional mutual information} of $X$ and $Y$ conditional on $Z$ as:
\eqn{
I(X:Y|Z) := H(XZ)+H(YZ)-H(XYZ)-H(Z) \, . \nonumber
}
Intuitively, this tells us how much the variables $X$ and $Y$ are correlated, conditional on the value of $Z$. The connection to CI relations is that the causal constraint $(X \LL Y|Z)$ is equivalent to the entropic constraint $I(X:Y|Z)=0$. 

Unless otherwise specified, the results in this work apply to any entropy function that satisfies the \tit{polymatroidal axioms} (also called \tit{elementary inequalities}, these are standard in information theory -- see eg. Ref. \cite{YEUNG}). Let $V$ denote a set of variables, $X \subseteq V$ an arbitrary subset, and $A,B$ specific variables in $V$. Then these inequalities are:
\eqn{ \label{eqn:element}
H(V \setminus A) &\leq& H(V) \, , \, \, \, \, \forall A \in V \, ;\nonumber \\
H(X)+H(X A B) &\leq& H(XA)+H(X B) \, , \, \, \, \, \forall (A \neq B) \in V \setminus X ;\nonumber \\
H(\emptyset) &=& 0 \, . \nonumber \\
&&
}

Any inequality that follows from these is called a \tit{Shannon-type} inequality. Given a set of variables $V$ distributed by $P(V)$, an entropy function assigns a real value $\geq 0$ to all subsets of $V$. If we regard each subset as the basis of an abstract vector space, then the entropy function maps the distribution $P$ to a vector in this space; the set of all such vectors (corresponding to valid probability distributions) defines a convex cone. An explicit characterization of this cone has not been found to date. However, since all valid entropy functions must satisfy the inequalities \eqref{eqn:element}, these define an outer approximation called the \tit{Shannon cone}. The entropy vector of any valid distribution $P(V)$ must lie within this cone, but the converse is not true -- there are sometimes distributions that satisfy all Shannon-type inequalities but cannot be obtained as the marginal of a valid probability distribution on $V$.

We are now in a position to formally state the problem of whether or not a GDAG is \tit{interesting}. Let $G(V,U)$ be a GDAG with observed variables $V$ and unobserved variables $U$. We are concerned with the question of whether all distributions $P(V)$ that respect the causal structure of $G$ can arise as the marginal distribution of a classical causal model. Following HLP, let $\mathcal{I}$ be the set of distributions $P(V)$ that respect the observable independencies implied by $G$, i.e. such that $\trm{CI}_{G}(V) \subseteq \trm{CI}_{P}(V)$. Next, define $\mathcal{C}$ as the set of distributions $P(V)$ that are marginals of some probability distribution $P(V,U)$ having $\trm{CI}_{G}(V,U) \subseteq \trm{CI}_{P}(V,U)$. Since the latter condition is strictly stronger, we have $\mathcal{C} \subseteq \mathcal{I}$, and the graph is called \tit{interesting} if and only if the inclusion is strict, $\mathcal{C} \neq \mathcal{I}$ (otherwise it is \tit{uninteresting}). Note that a difference between these sets indicates the existence of constraints on $\mathcal{C}$ that can be violated by distributions in $\mathcal{I}$. Thus, a sufficient condition to show that a GDAG is interesting is to identify a constraint in $\mathcal{C}$ that does not apply in $\mathcal{I}$. This can be established in two ways: (i) the constraint can be proven not to follow from the contraints that define the set $\mathcal{I}$, or (ii) the constraint can be shown to be violated by a specific distribution in $\mathcal{I}$. The first method is simplest (when it applies) and can be done by purely graphical techniques, some of which are reviewed in the next section. The second method requires searching the probability space of $\mathcal{I}$ for suitable counterexamples, but can succeed in many cases where the first method fails; in Sec \ref{Sec:newstuff} we employ this method using what we call `fine grained' inequalities. 

\section{The skeleton method and e-separation \label{Sec:skelly}}

In this section, we review two graphical techniques proposed by Evans for identifying cases where $\mathcal{C} \subsetneq \mathcal{I}$, that is, for identifying graphs as interesting. The first technique is called the \tit{Skeleton method}, and the second is called \tit{e-separation}. As we will see, these are not independent; the Skeleton method derives from a corollary of e-separation and is easier to understand and use. For graphs larger than 7 nodes, however, we argue that e-separation is the more general and more useful method.

We begin with some useful definitions adapted from Refs. \cite{EVANS12,EVANS14}. Where our terminology differs from that of Evans, a translation is provided in the footnotes. In what follows, $G(V,U)$ is a GDAG with observed variables $V$ and unobserved variables $U$. \\

\begin{figure}[!htbp]
\includegraphics[width=8cm]{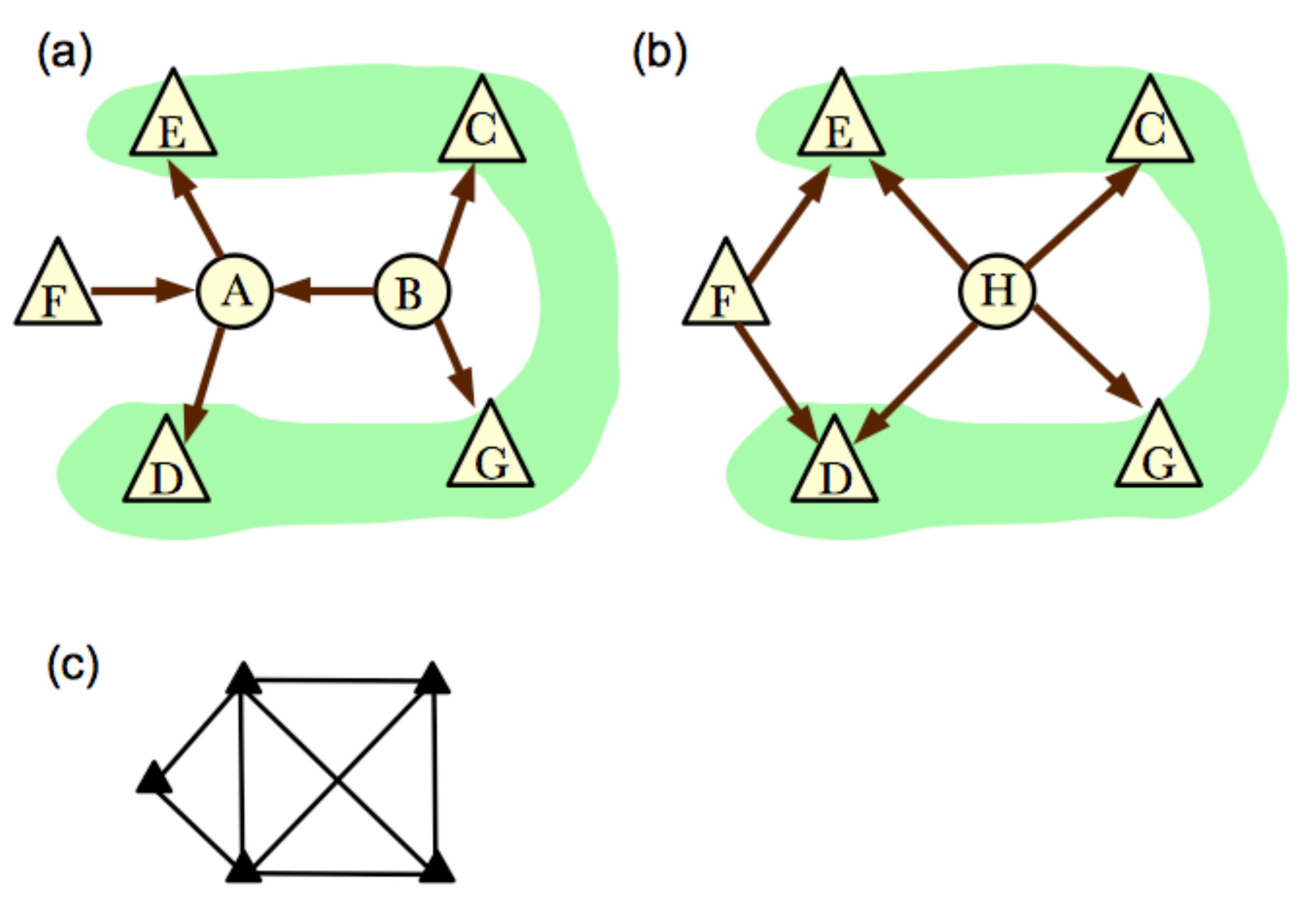}
\caption{Canonical projection of a GDAG. (a) shows the original GDAG, whose only maximal subset of size $\geq 2$ is $\{ C,D,E,G \}$ (highlighted). (b) is the corresponding canonical GDAG. By definition, both have the same skeleton, shown in (c). This example is taken from Evans \cite{EVANS14}.} 
\label{fig:canon}
\end{figure}

\tbf{Definition \defn{def:hiddP} \ref{def:hiddP}: hidden paths and causes}.\\
Let $X,Y$ be any specific variables in $G$. A directed path from $X$ to $Y$ is called a \tit{hidden path} iff it contains more than one arrow and all nodes on the path besides $X,Y$ are in $U$ (i.e. unobserved). Two variables $X,Y$ are said to share a \tit{hidden common cause} iff they share an unobserved common ancestor $W \in U$ to which they are either directly connected, or connected by a hidden path. (Note: a variable is not considered to be an ancestor of itself). For example, in Fig. \ref{fig:canon} (a), the node $B$ is a hidden common cause of the nodes $E$ and $C$.\\ 

\tbf{Definition \defn{def:maxB} \ref{def:maxB}: maximal connected subsets} \footnote{Our \tit{connected subsets} are Evans' \tit{bidirected faces}. Our \tit{maximal connected subsets} are his \tit{bidirected facets}. Our set $\scr{B}_G(V)$ is Evans' set $\bar{\scr{B}}$. Compare to Evans, \cite{EVANS14}.}.\\
In the GDAG $G(V,U)$, a subset of observed variables $B \in V$ is called \tit{connected} iff all nodes in $B$ share a single hidden common cause. A connected set $B$ is called \tit{maximal} iff there is no other connected set that fully contains it. Let $\scr{B}_G(V)$ denote the set of all maximal connected subsets (of size $\geq 2$) in $G$. For example, in Fig. \ref{fig:canon} (a), the sets $\{E,C\}, \, \{C, G\}$ are connected subsets, and $\{C,D,E,G\}$ is the only non-trivial maximal connected subset.\\

\tbf{Definition \defn{def:canon} \ref{def:canon}: canonical GDAG} \footnote{Our procedure of \tit{canonical projection} is, in Evan's language, equivalent to taking the \tit{canonical DAG} of the \tit{latent projection} over the unobserved nodes of the original GDAG.} .\\
Given the GDAG $G(V,U)$, define its \tit{canonical projection} $G'(V,W)$ as follows:\\
(i) The observed nodes of $G'$ are the same as those in $G$.\\
(ii) Two observed nodes $X,Y$ are connected by an arrow $X \rightarrow Y$ in $G'$ iff they are connected by an arrow or a hidden path in $G$.\\
(iii) There is one unobserved node in $W$ for each maximal connected subset of $G$, and each node in $W$ is a direct cause of all nodes in the corresponding element of $B \in \scr{B}_G(V)$. \\
Fig. \ref{fig:canon} shows an example of the canonical projection of a GDAG. A GDAG is said to be \tit{canonical} iff it is equal to its own canonical projection.\\

As shown by Evans (see Proposition 4.9(b) and Theorem 4.13 in Ref. \cite{EVANS14}), every GDAG is \tit{observationally equivalent} to its canonical projection, in the sense that they have the same set of possible marginal distributions $P(V)$ on the observed variables.\\ 

\tbf{Definition \defn{def:skel} \ref{def:skel}: GDAG skeleton}.\\
Let $G(V,U)$ be a canonical GDAG with maximal connected subsets $\scr{B}$. The \tit{skeleton} of $G$ is defined to be an undirected graph having nodes $V$, and where two nodes are connected iff they are connected by an arrow in $G$, or belong to the same maximal connected subset in $\scr{B}$.\\ 

\begin{figure}[!htbp]
\includegraphics[width=8cm]{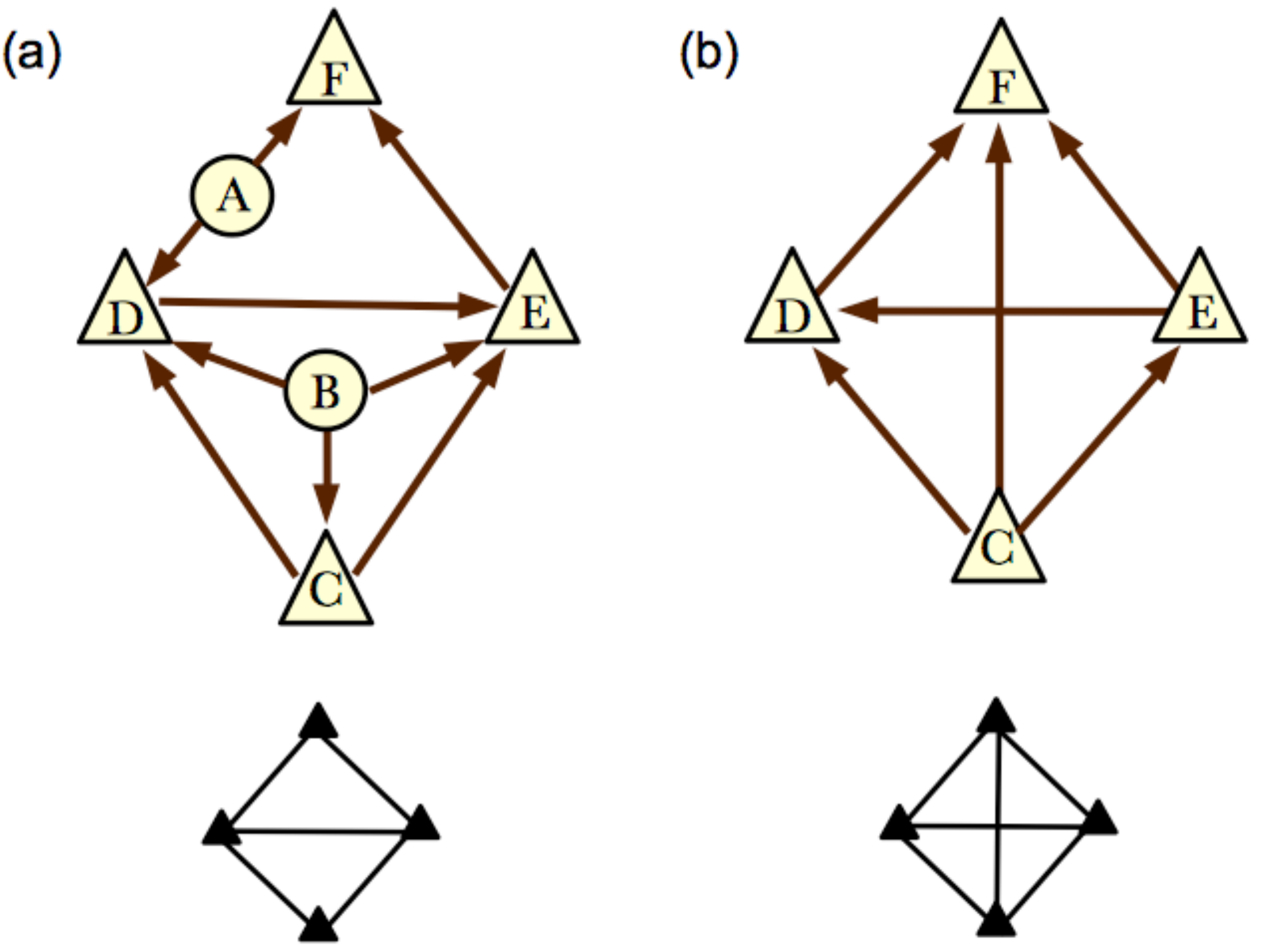}
\caption{Application of the Skeleton method to graph \#21 from HLP. The original GDAG and its skeleton are shown in (a). Since the graph has no observable conditional independencies, a candidate for the comparison graph $K$ is a complete graph, like the one shown in (b). Since its skeleton is different, Theorem \ref{thrm:skelmeth} implies the original graph is interesting.}
\label{fig:skel21}
\end{figure}

(If $G$ is not canonical, we define its skeleton to be the skeleton of its associated canonical GDAG). For example, Fig. \ref{fig:skel21} (a,b) shows two GDAGs with their corresponding skeletons underneath. We can now introduce the Skeleton method:\\

\tbf{Theorem \thrm{thrm:skelmeth} \ref{thrm:skelmeth}: The Skeleton method}. \\
Consider a GDAG $G(V,U)$. Let $K(V,W)$ be a GDAG with the same set of observable variables $V$ and observable CI relations $\trm{CI}_{G}(V)$, but for which it is known that $\mathcal{C} = \mathcal{I}$ (assuming such a GDAG can be found). Suppose furthermore that the skeletons of $G$ and $K$ are different. Then $\scr{C} \subsetneq \scr{I}$ for $G$.\\

The proof follows from Proposition 6.5 in Ref. \cite{EVANS14}, which states that the observed marginals of $G$ and $K$ are different if they have different skeletons. Since they have the same set of observable variables and observable CI relations, the set $\mathcal{I}$ is the same for both graphs, and since $\mathcal{C} = \mathcal{I}$ for $K$, it follows that $\scr{C} \subsetneq \scr{I}$ for $G$. $\Box$ \\

This method is useful for GDAGs having few or no observed CI relations, where a suitable GDAG $K$ is easy to find. For instance, if $\trm{CI}_{G}(V)=\emptyset$, one can always take $K(V)$ to be a maximally connected DAG. It is likely to become increasingly difficult to find a suitable graph $K$ with which to apply the Skeleton method for large numbers of variables; nevertheless the method works well for small GDAGs. Fig. \ref{fig:skel21} shows how it can be applied to graph \#21 from Ref. \cite{HLP}, and the reader is encouraged to verify that the same method can be used to establish the interestingness of most of the other graphs considered by those authors. The notable exceptions are \#2 (the Bell scenario), \#8 (the Triangle scenario), and \#15,16,20 which we consider in Sec. \ref{Sec:newstuff}. For the Bell scenario, the method fails because there is no graph $K$ that satisfies the no-signalling constraint and has $\scr{C} = \scr{I}$, while for the Triangle scenario it fails because the skeleton of all candidate $K$ graphs is the same as that of $G$. For \#15,16,20, the situation is analagous to the Bell scenario: for each one, there is no graph $K$ that satisfies the same observed CI relations yet has a different skeleton. \\

We now discuss a second graphical method based on the idea of `extended d-separation', or \tit{e-separation} in Evans \cite{EVANS12}. \\

\tbf{Definition \defn{def:esep} \ref{def:esep}: e-separation}.\\
Let $X,Y,Z,W$ be four disjoint sets of variables in a DAG. Suppose that, after removing the nodes $W$ from the graph, the sets $X$ and $Y$ are d-separated by the set $Z$ in the new graph. Then we say that `\tit{X and Y are e-separated by Z after deletion of W}'. \\

In the above definition, the new graph produced `after removal of the nodes $W$' is to be understood formally as the induced subgraph of the original graph on the complement of $W$, that is, the graph consisting of the nodes not in $W$, and the edges connecting these nodes that do not have endpoints in $W$. 

Notice that e-separation is a relation between four sets of nodes, and it reduces to the three-set criterion of d-separation when $W = \emptyset$. Evans showed that, under certain conditions, e-separation implies an extra constraint on the set of observed classical distributions $\scr{C}$. To explain this constraint, we need one more definition:\\

\tbf{Definition \defn{def:compat} \ref{def:compat}: compatibility along a section}.\\
Let $P(W,X,...)$ and $P^*(W,X,...)$ be two different probability distributions on the same set of variables. If they take the same values whenever the variables $W$ are fixed to a particular set of values $W=w$ (i.e. if $P(W,X,...) \, \delta(W,w) = P^{*}(W,X,...) \, \delta(W,w)$) then we say that $P$ and $P^{*}$ \tit{are compatible along the section W=w}.\\ 

We can now describe the constraint that is implied by e-separation:\\

\tbf{Theorem \defn{thrm:esepconst} \ref{thrm:esepconst}: e-separation constraint}.\\
Let $P(X,Y,Z,W,...)$ be any distribution in $\scr{C}$ for a DAG $G$. Suppose that $X$ and $Y$ are e-separated by $Z$ after deletion of $W$, and that no member of $Z$ is descended from $W$. Then for each possible value $w$ of $W$, the conditional distribution $P(X,Y,W,...|Z)$ must be compatible along the section $W=w$ with a distribution $P^{(w)}(X,Y,W,...|Z)$ in which $(X \LL Y | Z)$ holds.\\ 

The proof of this constraint on $\scr{C}$ is derived in Theorem 4.2 in Ref. \cite{EVANS12}. The question remains under what circumstances it represents a constraint on $\scr{I}$ as well. To answer this, we require the following Lemma:\\

\tbf{Lemma \thrm{thrm:lemma1} \ref{thrm:lemma1}: Conditions for e-separation to imply $\scr{C} \neq \scr{I}$}. \\
Let $G(U,V)$ be a GDAG in which $\{ X,Y,Z,W \} \subseteq V$ are disjoint subsets of observed nodes, such that no member of $Z$ is descended from $W$ and $X$ and $Y$ are e-separated by $Z$ after deletion of $W$. Then $\scr{C} \neq \scr{I}$ for $G$ if and only if the observed CI relations $\trm{CI}_{G}(V)$ exclude all relations of the form $(X \LL Y | Z S)$, where $S$ is any subset of $W$.\\

\tit{Proof:}
First, we prove the `only if' clause by showing that if any relation $(X \LL Y | Z S)$ does hold, then the e-separation constraint follows from the constraints $\trm{CI}_{G}(V)$ alone. Let $W = S \cup T$, and let $W$ take possible values $w=\{s,t \}$. Suppose that $(X \LL Y | Z S)$ holds in $\trm{CI}_{G}(V)$. For each $\{s,t \}$ we can then define $P^{(s,t)}(X,Y,Z,S,T,...) := P(X,Y,T,Z...| S) \, \delta(S,s)$. Since $(X \LL Y | Z S)$ necessarily holds in $P$, $(X \LL Y | Z)$ necessarily holds in $P^{(s,t)}(X,Y,S,T,...|Z)$. Moreover, each $P^{(s,t)}$ satisfies
\eqn{
&& P^{(s,t)}(X,Y,S,T,...|Z) \, \delta(S,s) \, \delta(T,t) = \\ \nonumber 
&& P(X,Y,T,...| Z S) \, \delta(S,s) \, \delta(T,t) \, ,
}
hence $P^{(s,t)}$ is compatible with $P$ along the section $ST = \{s,t\}$ after conditioning on $Z$. Therefore, the presence of $(X \LL Y | Z S)$ in $\trm{CI}_{G}$ is sufficient to imply the e-separation constraint of Theorem \ref{thrm:esepconst}.

Next, we prove the `if' clause by explicitly constructing a distribution $P'$ that satisfies $\trm{CI}_{G}(V)$ (and is therefore in $\scr{I}$) but cannot satisfy the e-separation constraint. Consider the distribution $P'(X,Y,Z,S,T...)$ in which $X,Y$ are perfectly correlated (and take more than one value) and every other variable is fixed to a single value. (Note that this distribution is only possible if $\trm{CI}_{G}(V)$ excludes all relations of the form $(X \LL Y | Z S)$, otherwise it is not a valid distribution for the GDAG $G$). We then find that there is only a single pair of values $\{s,t \}$ for which $P'$ is non-vanishing, which means that $P'$ is equal to its own ``section'', $P' = P' \, \delta(S,s) \, \delta(T,t) $. Moreover, the only normalized probability distribution that is compatible with $P'$ along this section is $P'$ itself, hence we must have $P^{(s,t)} := P'$. But this means $X,Y$ are necessarily correlated in $P^{(s,t)}$ conditional on $Z$, so the e-separation constraint is impossible to satisfy for the distribution $P'$. $\Box$ \\

These observations mean that, for a given graph, we can tell whether the e-separation constraint holds in $\scr{I}$ or not, and hence whether a graph is interesting:\\

\tbf{Theorem \thrm{thrm:esep} \ref{thrm:esep} : The e-separation method}. \\
Consider a GDAG $G$ with observed CI relations $\trm{CI}_{G}(V)$. Let $X,Y,Z,W$ be disjoint sets of observable nodes, such that no member of $Z$ is descended from $W$ in the graph. Then if $X$ and $Y$ are e-separated by $Z$ after deletion of $W$, and if $\trm{CI}_{G}(V)$ excludes $(X \LL Y|ZS)$ for all subsets $S \subseteq W$, then $\scr{C} \subsetneq \scr{I}$ for $G$. \\

The proof follows from Theorem 4.2 in Ref. \cite{EVANS12}, and our Lemma \ref{thrm:lemma1}. Note that in the cases where $\scr{C} \subsetneq \scr{I}$, the Lemma also tells us how to explicitly construct a distribution that violates the e-separation constraint: simply have $X,Y$ perfectly correlated and all other variables fixed. \\

\begin{figure}[!htbp]
\includegraphics[width=8cm]{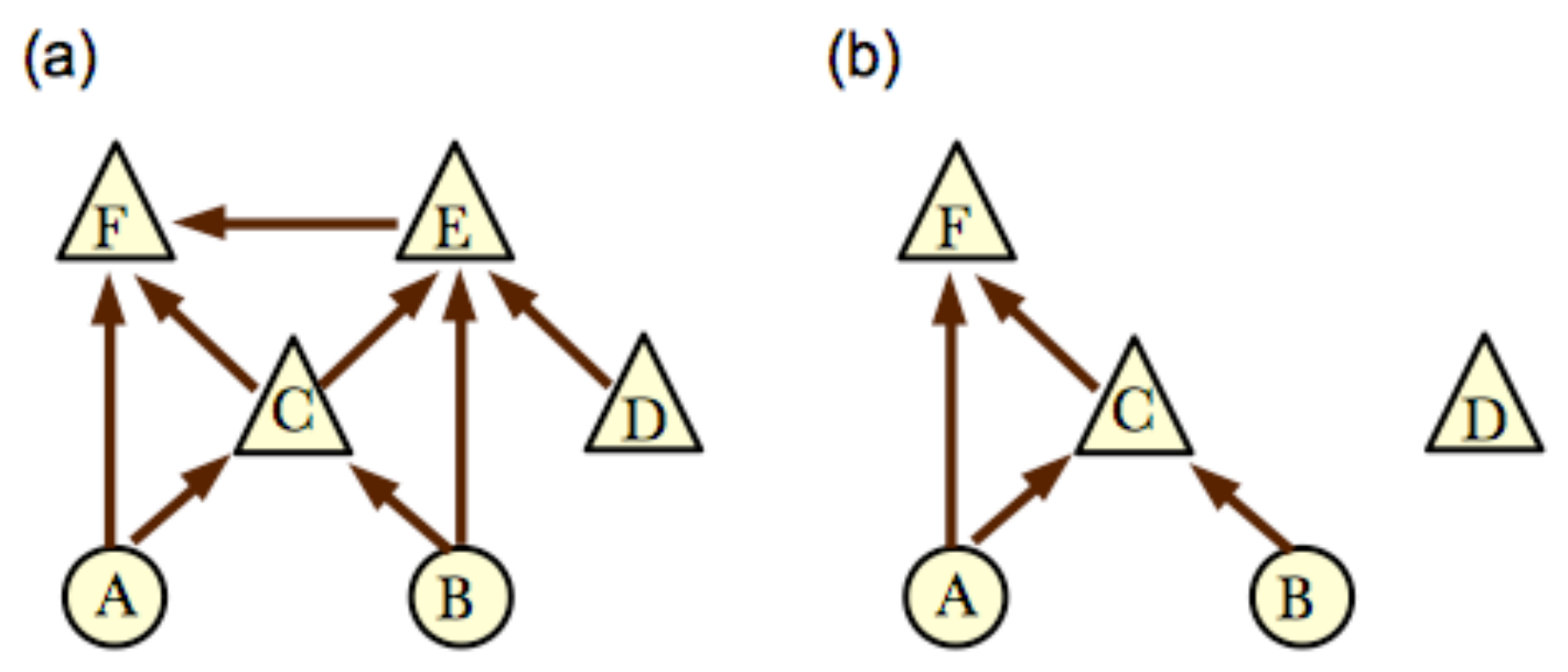}
\caption{Application of the E-separation method to graph \#17 from HLP. The original GDAG is shown in (a), while (b) shows the GDAG after deleting $E$ and its edges. Note that $C$ is not descended from $E$; also the CI relation $(F \LL D|C)$ holds in (b) but neither it nor $(F \LL D|CE)$ holds in $\trm{CI}_{G}$. Theorem \ref{thrm:esep} can then be applied to show the graph is interesting, if we identify $\{ F,D,C,E \}$ respectively with the sets $\{ X,Y,Z,W \}$ in the Theorem. } 
\label{fig:esep}
\end{figure}

Fig. \ref{fig:esep} shows how to apply the e-separation method to GDAG \# 17. In this graph, $F,D,C,E$ correspond respectively to the sets $\{ X,Y,Z,W \}$ in Theorem \ref{thrm:esep}, and $F$ and $D$ are e-separated by $C$ after deletion of $E$. This means $(F \LL D|C)$ holds in the graph after deletion, Fig. \ref{fig:esep} (b), but notice that neither $(F \LL D|C)$ nor $(F \LL D|CE)$ holds in the original graph Fig. \ref{fig:esep} (a). (To see this, observe that the path $F \leftarrow E \leftarrow D$ is unblocked if we condition only on $C$, but if we condition on both $C$ and $E$, then another path $F \leftarrow A \rightarrow C \leftarrow B \rightarrow E \leftarrow D$ becomes unblocked). Hence the conditions of Theorem \ref{thrm:esep} are met and the graph must be interesting. 

Note that instead of $(F \LL D|C)$, we could alternatively have used the CI relation $(F \LL D)$, i.e. taking the conditioned set $Z$ to be the empty set. On the other hand, we could \tit{not} have deleted $C$ and used the resulting relation $(F \LL D|E)$ to invoke E-separation, because $E$ is descended from $C$, violating a premise of Theorem \ref{thrm:esep}. \\

The e-separation method is based on Theorem 4.2 in Ref. \cite{EVANS12}, of which Proposition 6.5 in Ref. \cite{EVANS14} (the basis for the Skeleton method) is a Corollary. We therefore expect the e-separation method to succeed in at least those cases where the Skeleton method succeeded (and possibly more). In the Appendix, we present a GDAG of eight nodes whose interestingness can be proven by e-separation, but cannot be proven by the Skeleton method, thereby demonstrating that e-separation is a strictly more useful criterion.

For GDAGs of up to six nodes, we find the e-separation method has the same success as the Skeleton method: the reader is encouraged to verify that it confirms the interestingness of all GDAGs considered by HLP except the Bell scenario, the Triangle scenario, and the three GDAGs \#15,16,20. 

It fails for the Bell scenario because there is no way to delete any observed nodes to create a CI relation that is not already implied by $\trm{CI}_{G}(V)$. The Triangle scenario has a similar problem: to avoid triviality, one can only delete a single node, but then the remaining two nodes will still be correlated, i.e. no new observable CI relations are implied. The method fails for graphs \# 15,16,20, because the only viable candidates for the sets $X,Y$ in Theorem \ref{thrm:esep} either cannot be d-separated by the deletion of other observed nodes (as in \# 16), or the deletion of any candidate node does not result in new CI relations (as in \# 15, 20). 

\tit{Remark 1:} In the cases of up to six nodes where the technique does work, the distribution that violates the e-separation constraint is the same distribution that HLP found to violate a relevant Shannon-type entropic inequality. This indicates that whenever the e-separation constraint applies, there may exist a corresponding Shannon-type entropic inequality.

\tit{Remark 2:} In the Appendix we present a graph of 8 nodes that e-separation identifies as interesting, but where the Skeleton method fails to apply. This indicates that, for larger DAGs, e-separation is strictly more powerful than the Skeleton method.

\section{Entropic inequalities via `fine-graining' \label{Sec:newstuff}}

\begin{figure}[!htbp]
\includegraphics[width=9cm]{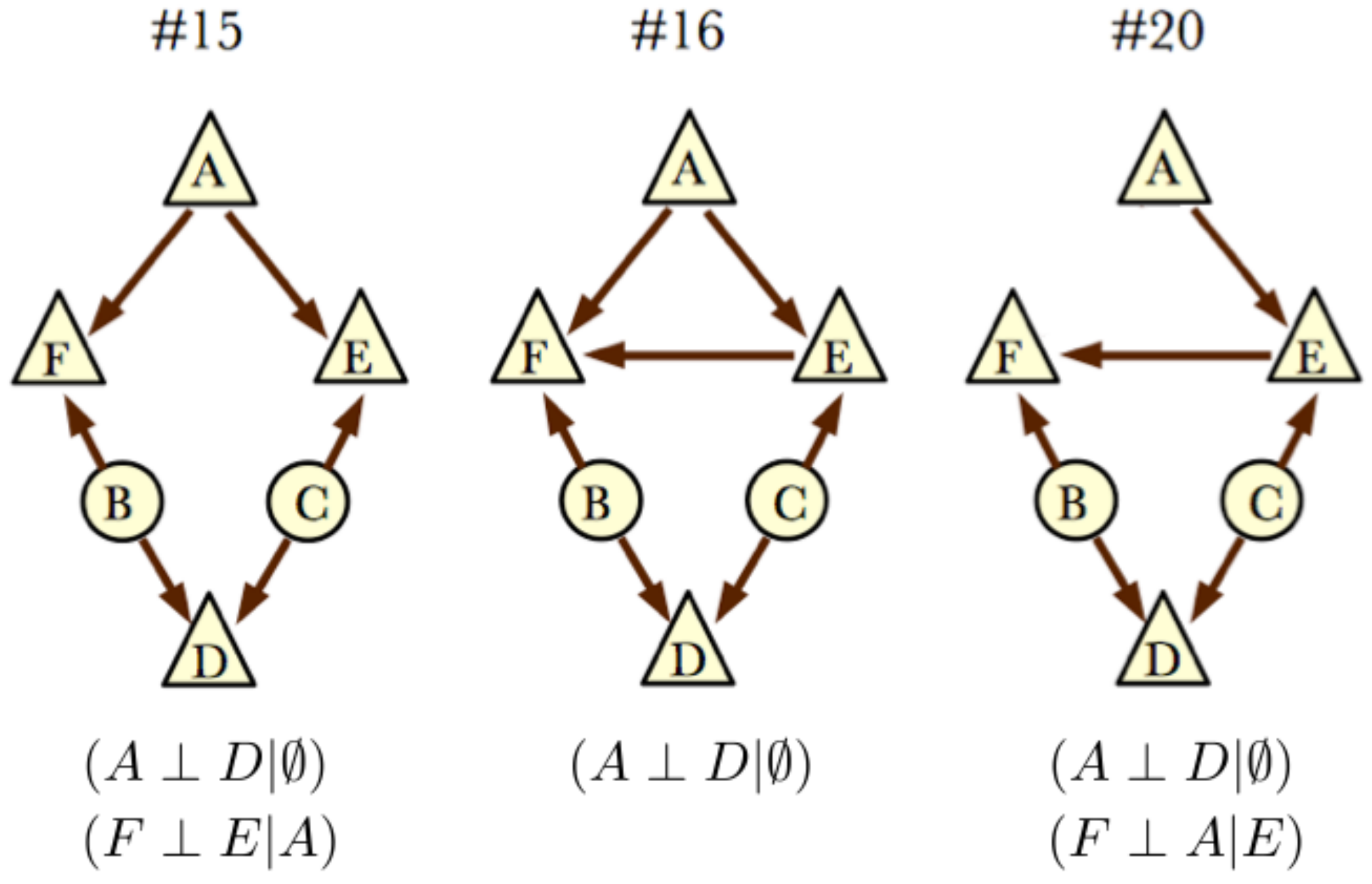}
\caption{The causal graphs \#15,16,20 from HLP, with observed CI relations written below. These do not satisfy HLPs' sufficient condition for uninterestingness, but they also resist Evans' sufficient criteria for interestingness. We find a probability distribution that violates the `fine-grained' entropic inequality \eqref{eqn1}, thereby confirming that these graphs are indeed interesting.}. 
\label{fig:musketeers}
\end{figure}

We now turn to the problem of establishing the interestingness of the graphs \# 15,16,20 as conjectured in HLP (these are displayed in Fig. \ref{fig:musketeers}). 

Let us concentrate on graph $\# 16$; the others will follow in kind. For simplicity, we assume $A$ is binary and takes values in $\{ 0,1 \}$. Our goal is to demonstrate that the observed marginal of this graph is subject to an entropic constraint, namely:
 
\eqn{ \label{eqn1}
&& I(E:D|\hat{A}_0)-I(F:D|\hat{A}_0)  \,  \nonumber \\ 
 + && \, \, \, I(F:D|\hat{A}_1)-I(E:D|\hat{A}_1) \leq H(D) \, , \nonumber \\ 
&&
}

where the notation $\hat{A}_0$ means $\trm{\tbf{do}}(A=0)$, i.e. `intervene on $A$ and set it to $0$' . Thus, for example, $I(E:D|\hat{A}_0)$ is the mutual information of $E,D$ in the post-intervention distribution $P(E,D|\hat{A}_0)$. The reader might object that, if we cannot perform interventions, then we cannot obtain the quantities appearing on the left hand side of the inequality \eqref{eqn1}. In fact, we \tit{can} obtain these quantities without doing an intervention, because $A$ is an exogenous node in the graph (it has no parents). Hence the following identity holds:
\eqn{ \label{docalc1}
P(E,D|\hat{A}_{0})=P(E,D|A)\, \delta_{(A,0)} \, ,
}
where $\delta_{(A,0)}$ is a Kronecker delta function. Intuitively, this says that the post-intervention distribution is the same as the distribution obtained by \tit{post-selecting} on $A$ having the particular value $0$ or $1$. We therefore have a situation where the causal graph allows us to learn something about the post-intervention distributions without actually performing an intervention. Whenever a constraint such as the inequality \eqref{eqn1} involves these post-intervention distributions, we call it a \tit{fine-grained} constraint. We now establish a general framework for obtaining fine-grained entropic constraints from a causal graph, and use it to derive Eqn. \eqref{eqn1}.

\subsection{Preliminaries \label{prelim}}

Let $P(R,X,Z)$ be a classical distribution (all variables are observed) compatible with some DAG $G(R,X,Z)$ in the usual sense that $\trm{CI}_G(R,X,Z) \subseteq \trm{CI}_P(R,X,Z)$. Here, $X,Z$ are specific variables, $R$ is the set of remaining variables, and $Z$ is exogenous. Then the distribution factorises as $P(R,X,Z)=P(R,X|Z) P(Z)$. In the following we assume $Z$ is binary for clarity, but the results can easily be extended to the case where $Z$ is an arbitrary discrete variable. We give definitions for the case $Z=0$; similar definitions are understood to hold for $Z=1$.\\

Define the notation:
\eqn{ \label{Zdefin0}
P(R,X|\hat{Z}_{0}) &:=& P(R,X|\trm{\tbf{do}}(Z=0)) .
}

Therefore,

\eqn{ \label{Zdefin}
P(R,X|\hat{Z}_{0}) &=& P(R,X|Z)\delta_{(Z,0)} \, .
}

For any subset of variables $S$ (excluding $Z$), we define $H(S|\hat{Z}_{0})$ as the entropy of $S$ with respect to the post-intervention distribution $P(S|\hat{Z}_{0})$:
\eqn{
H(S|\hat{Z}_{0}) := - \sum_{S} \, P(S|\hat{Z}_{0}) \trm{log}_2 \left[ P(S|\hat{Z}_{0}) \right] \, .
} 
Entropies of this form (pertaining to a specific value of $Z$) are called `fine-grained'. Functions of the entropies, like the mutual information $I(X : Y|\hat{Z}_{0})$, are defined analogously, using the entropies of the post-intervention distribution. For example:
\eqn{
I(X : Y|\hat{Z}_{0}) &:=& H(X|\hat{Z}_{0})+H(Y|\hat{Z}_{0})-H(XY|\hat{Z}_{0}) \, . \nonumber \\
&&
}
Any inequality that contains terms like these is said to be `fine-grained'. Any CI relations that hold in the original distribution necessarily hold in the post-intervention distribution. For example, if $I(X : Y|Z)=0$ holds in $P(X,Y,Z)$, then the statement $I(X : Y|\hat{Z}_{0})=0$ is also valid. Moreover, we have: \\

\tbf{Corollary \thrm{thrm:docalc} \ref{thrm:docalc} :} \\
If $(S \LL Z)$ holds in $P(S,Z)$, then $P(S|\hat{Z}_{0})=P(S|\hat{Z}_{1})=P(S)$.\\

This statement is a consequence of Rule 3 in Pearl's \tit{do calculus} (see Pearl \cite{PEARL}, Theorem 3.4.1). It is important because it allows us to relate the entropies of distributions for which $Z=0$ to entropies of distributions in which $Z=1$, which implies fine-grained entropic constraints. In what follows, we only make use of Corollary \ref{thrm:docalc}, but in general the entire \tit{do calculus} is available for deriving such constraints. We return to this point in Sec. \ref{discus}.

\subsection{Derivation of the inequality \label{deriv}}

Returning to the specific case of graph $\# 16$, we aim to prove Eqn. \eqref{eqn1}. To this end we require the following elementary entropic inequalities:
\eqn{
&& H(DE) \leq H(E) +H(D) \, , \label{dea} \\
&& H(E)+H(CDE) \leq H(CE) +H(DE) \, , \label{cdea} \\
&& H(DF) \leq H(F) +H(D) \, , \label{dfa} \\
&& H(F)+H(BDF) \leq H(BF) +H(DF) \, , \label{bdfa} \\
&& H(D) + H(BCD) \leq H(CD)+H(BD)  \, , \label{bcd} 
}
as well as the CI relations:
\eqn{
&& (D \LL E|CA) \, ,  \label{ciCA} \\
&& (D \LL F|BA) \, ,  \label{ciBA} \\
&& (BCD \LL A) \, ,  \label{ciA} \\
&& (B \LL C)\, ,  \label{ciBC}
}
which happen to hold in all three graphs. It is useful to define the quantities:
\eqn{
Q_0 := H(E|\hat{A}_{0})-H(DE|\hat{A}_{0})  \, ; \nonumber \\
R_0 := H(F|\hat{A}_{0})-H(DF|\hat{A}_{0})  \, ; \nonumber \\
Q_1 := H(E|\hat{A}_{1})-H(DE|\hat{A}_{1})  \, ; \nonumber \\
R_1 := H(F|\hat{A}_{1})-H(DF|\hat{A}_{1})  \, , \nonumber 
} 
so that the LHS of \eqref{eqn1} can be written (after expanding and cancelling some terms) as:

\eqn{ \label{eqnLHS}
\trm{LHS}  = \, \, \, Q_0 - R_0 +R_1-Q_1 \, . \,
}

From \eqref{cdea} \& \eqref{ciCA} we obtain:
\eqn{
&& H(E|\hat{A}_{0})-H(DE|\hat{A}_{0}) \leq H(C|\hat{A}_{0})-H(CD|\hat{A}_{0}) \,  , \nonumber \\
&&
}
and from \eqref{dea} we obtain:
\eqn{
-H(D|\hat{A}_{1}) \leq  H(E|\hat{A}_{1})-H(DE|\hat{A}_{1}) \, .
}
Similarly, from \eqref{bdfa} \& \eqref{ciBA}, and from \eqref{dfa}  we obtain:
\eqn{
 H(F|\hat{A}_{1})-H(FD|\hat{A}_{1}) &\leq& H(B|\hat{A}_{1})-H(BD|\hat{A}_{1}) \nonumber \\
-H(D|\hat{A}_{0}) &\leq& H(F|\hat{A}_{0})-H(FD|\hat{A}_{0})  \,  . \nonumber \\
&&
}
It follows that:
\eqn{
 Q_0 &\leq& H(C|\hat{A}_{0})-H(CD|\hat{A}_{0})  \, ; \nonumber \\
 - R_0 &\leq& H(D|\hat{A}_{0})  \, ; \nonumber \\
 R_1 &\leq& H(B|\hat{A}_{1})-H(BD|\hat{A}_{1})  \, ; \nonumber \\
 - Q_1 &\leq& H(D|\hat{A}_{1})  \, . \nonumber 
}
But $\eqref{ciA}$ implies:
\eqn{
&& H(D|\hat{A}_{0})=H(D|\hat{A}_{1})=H(D)  \, ; \nonumber \\
&& H(C|\hat{A}_{0})-H(CD|\hat{A}_{0}) = H(C)-H(DC) \, ; \nonumber \\
&& H(B|\hat{A}_{1})-H(BD|\hat{A}_{1}) = H(B)-H(DB) \, ,
}
so the upper bounds simplify to:
\eqn{
 Q_0 &\leq& H(C)-H(DC) \, ; \nonumber \\
 - R_0 &\leq& H(D)   \, ; \nonumber \\
 R_1 &\leq& H(B)-H(DB)  \, ; \nonumber \\
 - Q_1 &\leq& H(D)   \, . \nonumber 
}
Combining these with \eqref{eqnLHS}, we find the LHS of \eqref{eqn1} is upper bounded by:
\eqn{
\trm{LHS} &\leq& H(B)+H(C)-H(CD)-H(BD) +2 H(D) \, . \nonumber \\
&&
}
Finally, using \eqref{bcd} and \eqref{ciBC}, this simplifies to:
\eqn{
\trm{LHS} &\leq&  H(BC)-H(BCD)+H(D) \nonumber \\
&\leq& H(D) \, ,
}
where in the last step we used the elementary inequality $H(BC) \leq H(BCD)$. This completes the proof of \eqref{eqn1}. $\Box$ \\

It follows that the entropic inequality \eqref{eqn1} holds in graph $\# 16$; in fact, since the CI relations used in the proof are common to all three DAGs, the same inequality can be shown to hold in all three. Furthermore, we now present a probability distribution on the observed variables that satisfies all the observed CI relations of both $\# 15, \# 16$ but violates this inequality. Let $\tilde{P}(A,D,E,F)$ be an observed marginal distribution having the special property that when $A=0$, the bits $E,D$ are perfectly correlated and $F$ is fixed to $0$, whereas when $A=1$ the bits $F,D$ are correlated and $E$ is fixed. More formally, $\tilde{P}$ satisfies:\\

\noindent (i) all variables are binary with values in $\{ 0, 1 \}$; \\
(ii) the bits $A$ and $D$ are random and evenly distributed; \\
(iii) $F= A \times D$ ; \\
(iv) $E=(A \oplus 1)  D$ , \\

\noindent where $\oplus$ is addition modulo 2, and $A \times D$ means the product of the values of $A,D$.
Although $\tilde{P}$ satisfies all of the CI relations required by the DAGs $\# 15, 16$ for the \tit{observed} variables, it violates the constraint Eqn. \eqref{eqn1}, since (using the Shannon entropy) the LHS evaluates to 2, whereas the bound $H(D)$ for a binary variable $D$ is only 1. Hence $\tilde{P}(A,D,E,F)$ cannot be a (classical) marginal of the underlying DAG. The existence of this distribution therefore establishes $\scr{C} \subsetneq \scr{I}$ for $\#15,16$.

The case of $\# 20$ requires a little more care, because the distribution $\tilde{P}$ does not satisfy the relation $(F \LL A | E)$, which holds in $\# 20$. Fortunately, a minor adjustment of $\tilde{P}$ will allow us to obtain a distribution which does the job. Consider a distribution in which we allow $E$ to take \tit{three} possible values, $E \in \{0,1,2 \}$, but all other variables remain binary. Define the distribution $\tilde{P}'$ by the following property: when $A=0$, the bits $E,D$ are perfectly correlated and have values in $\{0,1\}$, while $F$ is fixed to $0$; when $A=1$ the bits $F,D$ are perfectly correlated and $E$ is fixed to the value $2$. Formally:\\

\noindent (i)  $A,D,F \in \{ 0, 1 \}$, $E \in \{ 0, 1, 2 \}$ ; \\
(ii) the bits $A$ and $D$ are random and evenly distributed; \\
(iii) $F= A \times D$ ; \\
(iv) $E=(A \oplus 1)  D + 2 A$ , \\

(note that the second plus sign in (iv) is not modulo 2). This distribution is almost identical to the previous one, except that knowledge of the value of $E$ is now sufficient to deduce the value of $A$. This means the distribution additionally satisfies $(F \LL A | E)$, and therefore satisfies all the observed CI relations of $\# 20$. As before, this distribution achieves 2 on the LHS and 1 on the RHS of Eqn. \eqref{eqn1}, violating the inequality.

\tit{Remark:} One could of course have used $\tilde{P}'$ to violate the inequality for all three graphs (since it satisfies the CI relations of all three), eliminating the need for $\tilde{P}$; however, we have chosen to include the latter distribution in our discussion because, while its scope is more limited, it is simpler in that it uses only binary variables.

In conclusion, the above distributions $\tilde{P},\tilde{P}'$ satisfy the observed CI relations of $\#15,\#16$, and $\#20$, respectively, but violate the inequality Eqn. \eqref{eqn1} that holds in these graphs, establishing $\scr{C} \subsetneq \scr{I}$. This completes the missing piece of HLP's analysis of DAGs of up to six nodes.

\subsection{Discussion \label{discus}}

The previous section suggests a general approach to finding fine-grained entropic inequalities such as Eqn. \eqref{eqn1}. One first selects one or more exogenous nodes to be `fine-grained', which means that we specify the set of possible values of each of these variables (eg. that they are binary and take values in $\{ 0,1 \}$ ). We then consider the entropic inequalities conditioned on specific values of the fine-grained nodes (i.e. the fine-grained versions of the usual entropic inequalities). Corollary \ref{thrm:docalc}, together with the CI relations from the graph, implies constraints relating entropies that are fine-grained on different values of the same variable. These constraints take the form of inequalities like Eqn. \eqref{eqn1}. The usual methods for finding entropic inequalities apply - the only difference is that the set of inequalities is enlarged to include fine-grained entropies as well as normal entropies.

The method we employed in the previous section has the disadvantage that it still involves finding explicit inequalities and searching for a distribution that violates one of them. Fortunately, Wolfe et. al. recently proposed a convenient graphical method called the \tit{inflation DAG} technique, with which they were independently able to establish the interestingness of GDAG $\#15$, as well as the Triangle and Bell scenarios\cite{WOLFE}.

The distribution $\tilde{P}$ described in the previous section is notable because it does not violate any standard Shannon-type entropic inequalities. In fact, the distribution lies inside an inner approximation of the entropy cone for scenario $\#15$ (the author thanks Weilenmann and Colbeck for this information \footnote{Weilenmann and Colbeck, private communication}). This means that no conventional entropic inequality can distinguish it from a valid distribution for this scenario. We conclude that fine-grained inequalities are strictly more powerful than conventional entropic inequalities. 

An interesting feature of graph $\# 15$ is that if we make $A$ into an unobserved variable, we recover the Triangle scenario. It turns out that the marginal obtained by summing over $A$ in $\tilde{P}$ is \tit{also} incompatible with the Triangle scenario \cite{WOLFE}. This suggests that although the ``W-distribution" discussed in Ref. \cite{WOLFE} lies inside an inner approximation of the entropic cone for the Triangle scenario, it might be possible to distinguish it from a valid distribution using a fine-grained entropic inequality. However, since all observed variables in that scenario have parents, this would require an extension of the present formalism to non-exogenous nodes.

From the presentation given here, there seem to be two obvious ways to extend the approach. The first method is to consider fine-graining by post-selecting on the values of non-exogenous nodes. For example, we could consider the distribution $P(R,X|Z=0)$ obtained from $P(R,X,Z)$ by post-selecting on the outcome $Z=0$. In the case where $Z$ has parents in the DAG, the resulting distribution is:
\eqn{
P(R,X|Z=0) = \frac{P(R,X,Z) \, \delta_{(Z,0)}}{P(Z=0)} \, ,
}
i.e. the distribution of $R,X$ conditional on $Z=0$. This can be thought of as a generalisation of the RHS of Eq. \eqref{Zdefin} because it reduces to the same expression in the case that $Z$ is exogenous. We could then search for fine-grained entropic inequalities conditioned on the values of non-exogenous nodes.

Note that, in general, the post-selected distribution $P(R,X|Z=0)$ is not the same as the post-intervention distribution $P(R,X|\trm{\tbf{do}}(Z=0))$. This brings us to the second possible extension, already hinted at earlier in the paper: to use post-intervention distributions like $P(R,X|\trm{\tbf{do}}(Z=0))$ in situations where $Z$ is not exogenous. One could then make use of the full \tit{do calculus} in order to discover new fine-grained entropic inequalities. We leave it to future work to explore the potential of these extensions of the fine-graining method.

\section{Conclusions \label{Sec:ending}}

The experimentally observed violation of Bell inequalities brings home the remarkable fact that, if we are certain that we know the causal structure of an experimental set-up, we can use it to make inferences about the fundamental laws of physics that govern the systems in the experiment. In particular, Bell-type experiments leverage the causal structure of space-time to rule out any faithful classical causal explanation of the experimental data. More generally, it would be useful to know which causal structures (represented by an associated GDAG) can be utilised for this purpose.

As pointed out in HLP, a necessary first step is the ability to identify the \tit{uninteresting} graphs, which cannot be used for distinguishing physical theories. Whereas HLP provided a sufficient condition for uninterestingness, in this paper we drew attention to some necessary conditions due to Evans (namely, we pointed out that a graph must be interesting if it violates Evans' graphical criteria reviewed in Sec. \ref{Sec:skelly} ). Furthermore, we derived a fine-grained entropic inequality that is able to confirm the interestingness of graphs \#15,16,20. Inequalities of this type were shown to follow from Pearl's \tit{do calculus} of interventions, suggesting the possibility of a general method for deriving such inequalities in more complicated scenarios. \\

\tit{Acknowledgements:} The author thanks Matt Pusey, Robin J. Evans, Elie Wolfe, Rafael Chaves and an anonymous referee for suggestions that greatly improved this paper. This work has been supported by the European Commission Project RAQUEL, the John Templeton Foundation, FQXi, and the Austrian Science Fund (FWF) through CoQuS, SFB FoQuS, and the Individual Project 2462.

%\bibliography{qdagRefs}{}
%\bibliographystyle{apsrev4-1}


\begin{appendix}*

\section{e-separation vs. the Skeleton method}

\begin{figure}[!htbp]
\includegraphics[width=8cm]{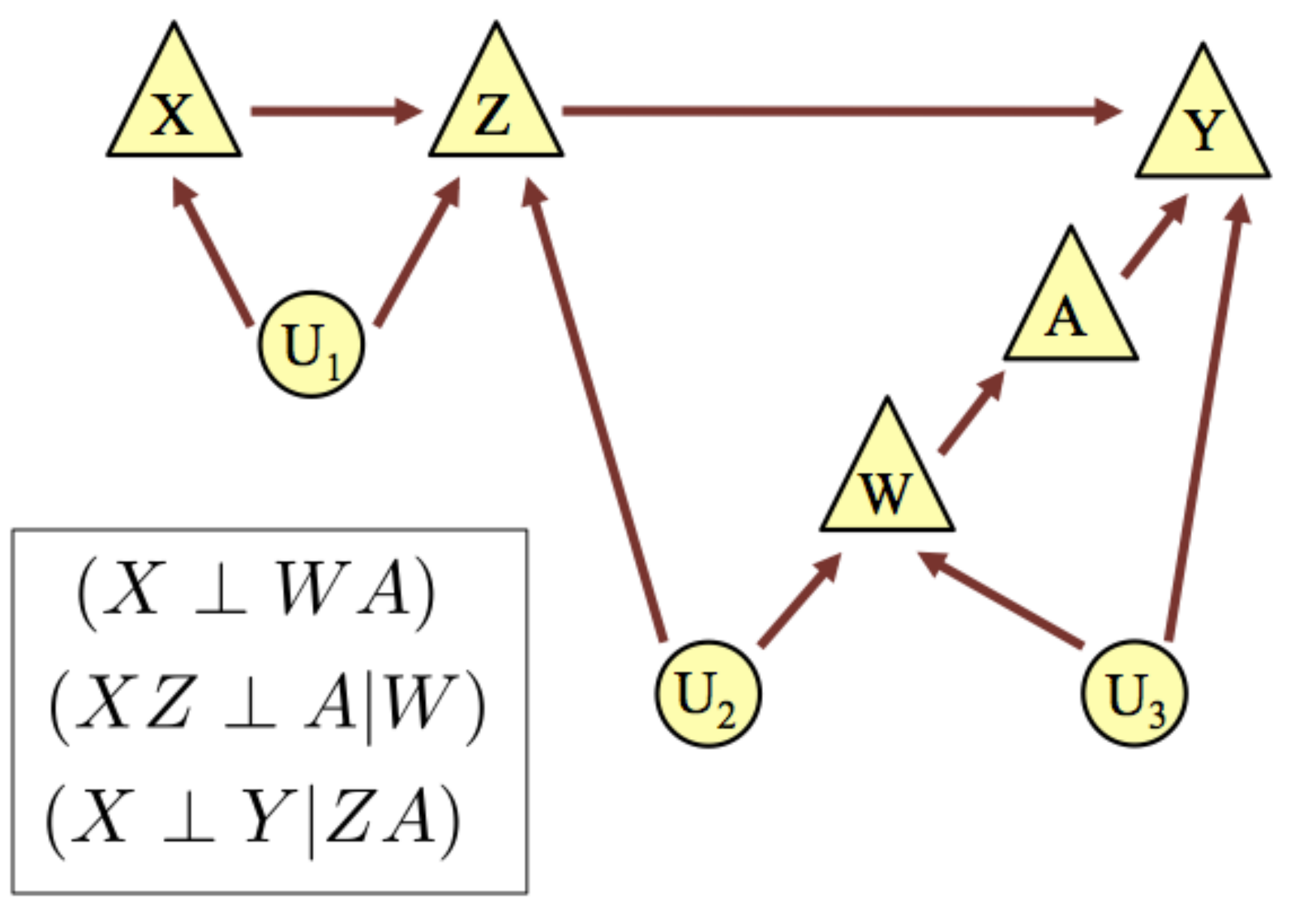}
\caption{A graph $G$ that is shown to be interesting via the e-separation method, but to which one cannot apply the Skeleton method. The CI relations in the box generate all relevant observed CI relations.} 
\label{fig:finalboss}
\end{figure}

\begin{figure}[!htbp]
\includegraphics[width=7cm]{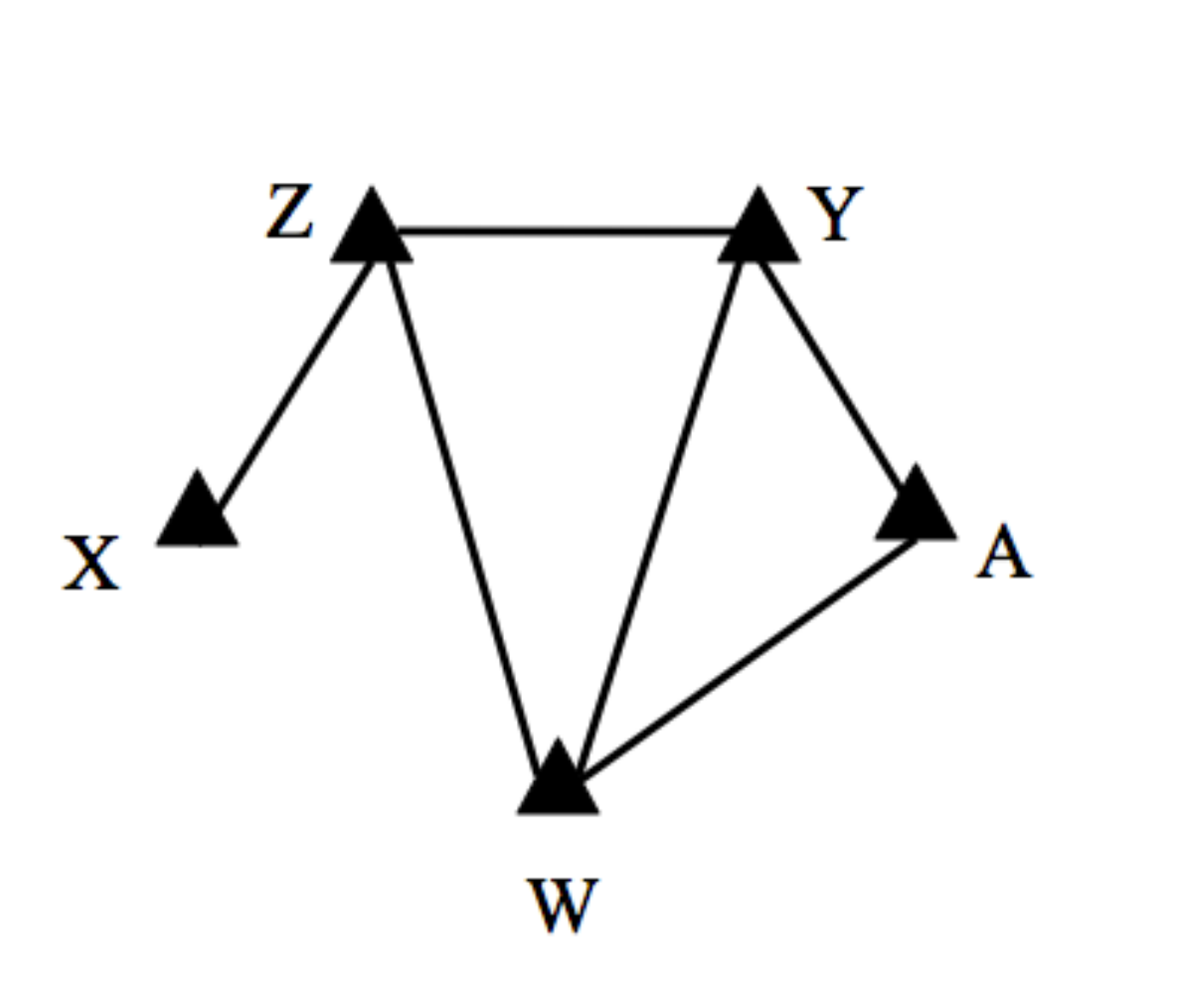}
\caption{The skeleton of the graph in Fig. \ref{fig:finalboss} .} 
\label{fig:bosskel}
\end{figure}

In Sec. \ref{Sec:skelly}, it was claimed that a GDAG exists in which e-separation could be used to determine $\scr{C} \neq \scr{I}$, but for which the Skeleton method could not be used. Here, we describe the GDAG and justify the claim. The GDAG is shown in Fig. \ref{fig:finalboss}, and its skeleton in Fig. \ref{fig:bosskel}. The three CI relations listed in Fig. \ref{fig:finalboss} are a generating set for the observed CI relations $\trm{CI}_{G}(V)$ where $V := \{X,Y,Z,W,A\}$. The variables $\{U_1, U_2, U_3\}$ are unobserved. 

To begin, let us determine that the graph $G$ is interesting, according to the criteria of e-separation. Following Theorem \ref{thrm:esep}, let the sets $X,Y,Z,W$ in the Theorem correspond to $X,Y,Z,W$ respectively in the graph. One can check that $X$ and $Y$ are e-separated by $Z$ after deletion of $W$; $Z$ is descended from $W$; and neither $(X \LL Y| Z)$ nor $(X \LL Y| Z W)$ is implied by $G$ (using the rules of d-separation). Thus, from Lemma \ref{thrm:lemma1} we have $\scr{C} \subseteq \scr{I}$ for $G$. 

Next, we show that the skeleton method cannot be used to reach the same conclusion. Specifically, we show that there is no GDAG $K$ whose skeleton is different to that of $G$ (as displayed in Fig. \ref{fig:bosskel}) and which satisfies the same observed CI relations as $G$. First, we observe that any candidate GDAG $K$ cannot have chords in its skeleton connecting the following pairs of nodes: $(X,A)$; $(X,W)$; $(X,Y)$; $(Z,A)$. This is because the presence of any one of these chords would violate an observed CI relation in all GDAGs consistent with the skeleton. For example, the chord $(X,Y)$ implies that any GDAG with that skeleton must have either a direct cause between $X,Y$ or else the two must share a hidden common cause; this means they cannot then be rendered independent by conditioning on other variables, violating $(X \LL Y|ZA)$. 
After eliminating these chords from the skeleton, the remaining chords are $(X,Z)$; $(Z,Y)$; $(Z,W)$; $(A,Y)$; $(A,W)$; $(Y,W)$. These chords are all present in the skeleton of $G$, shown in Fig. \ref{fig:bosskel}, hence any candidate GDAG $K$, to be useful for the Skeleton method, must possess only a strict subset of these chords in its skeleton.

We now show that removing \tit{any one} of the chords in $G$'s skeleton necessarily introduces a new CI relation not implied by $G$. To show this, we make use of the d-separation criterion for whether a path through the graph is `unblocked' or `blocked' (recall Definition \ref{def:dsep}). We treat each chord in turn.\\

\tit{Delete (X,Z)}: Removing this chord makes $X$ an isolated variable. Thus, eg. $(X\LL Z)$ must hold, but this is not implied by $G$. $\Box$ \\

\tit{Delete (Z,Y)}: Removing this chord leaves only one path between $X$ and $W$, namely $X - Z - W$. In order for the observed CI relation $(X \LL W)$ to hold, $Z$ must be a collider on this path. But then all paths connecting $X$ to $Y$ would be blocked, implying $(X \LL Y)$, which is not implied by $G$. $\Box$ \\

\tit{Delete (Z,W)}: Removing this chord leaves only one path connecting $X$ to $Y$, namely $X - Z - Y$. This path cannot be blocked by $A$. That means $(X \LL Y | Z A)$ only holds if $(X \LL Y | Z)$ holds -- but while the former relation is implied by $G$, the latter is not. $\Box$ \\

\tit{Delete (A,Y)}: Removing this chord means that no paths connecting $X$ and $Y$ contain $A$, so these paths cannot be blocked by $A$. Thus, as in the previous case, $(X \LL Y | Z A)$ requires $(X \LL Y | Z)$ to hold, but this is not the case in $G$. $\Box$ \\

\tit{Delete (A,W)}: Removing this chord again leaves no path between $X$ and $Y$ that contains $A$; thus, as in the previous two cases, $(X \LL Y | Z A)$ cannot hold without $(X \LL Y | Z)$, but only the former of the two is implied by $G$. $\Box$ \\

\tit{Delete (Y,W)}: This case is more complicated. First, it will be useful to establish that there is a directed path from $Z$ to $Y$ that does not contain any other observed nodes, and in addition $X,Y$ do not share a hidden common cause. We will indicate this by the shorthand $Z \rightarrow Y$ (abusing notation). To prove this, note that $(X \LL Y|A Z)$ (which is implied by $G$) means that all paths between $X$ and $Y$ are blocked by $Z,A$. Since the path $X-Z-Y$ only contains $Z$, this path must be blocked by $Z$. This can only occur if either $X \leftarrow Z$, or $Z \rightarrow Y$, or both. However, the former can't be the case, since then the path $X - Z - W$ would be unblocked, violating $(X \LL W)$, which is implied by $G$. So it must be the latter case: $Z \rightarrow Y$. \\
Now, if we remove the chord $(Y,W)$, there are only two paths connecting $Y$ and $W$, namely $W-Z-Y$ and $W-A-Y$. In order to avoid $(Y \LL W|AZ)$, which is not implied by $G$, we require that at least one of these paths is unblocked conditional on $AZ$, hence that $A$ and/or $Z$ must be a collider on its respective path. Since we have established $Z \rightarrow Y$, $Z$ cannot be a collider on the path $W-Z-Y$. That leaves $A$, but if $A$ is a collider on $W-A-Y$, then conditioning on $A$ cannot block any path between $X$ and $Y$. But then we are in the same position as the previous three cases: since $(X \LL Y|AZ)$ holds in $G$, so should $(X \LL Y|Z)$, but the latter is not implied by $G$. $\Box$ \\

We have seen that no single chord can be removed from the skeleton of $G$ without introducing new constraints not implied by $G$. However, since a CI relation cannot be invalidated by deleting further nodes or edges from a graph, it follows that the removal of \tit{any} set of chords from the skeleton of $G$ will result in new CI relations not implied by $G$. Together with the fact (proven above) that no chords can be added, we conclude that the skeleton of $G$ is the only one that supports the exact set of observed CI relations $\trm{CI}_{G}(V)$. Hence there is no GDAG $K$ that can be used for the purposes of the Skeleton method. The GDAG of Fig. \ref{fig:finalboss} is therefore a counterexample to the proposition that the power of e-separation is the same as the Skeleton method -- it shows that the former method is strictly more powerful in GDAGs of eight nodes or more. (Whether a counterexample exists with only seven nodes is an open problem, but I suspect the answer is negative).

\end{appendix}

\end{document}